\documentclass[aps,prl,preprint,groupedaddress]{revtex4}

\usepackage{graphics}
\usepackage{boxedminipage}
\usepackage{amsmath}
\usepackage{graphicx}% Include figure files
\usepackage{dcolumn}% Align table columns on decimal point
\usepackage{bm}% bold math

\begin{document}

\preprint{APS/unknown}

\title{Interferometric differentiation between resonant Coherent Anti-Stokes Raman Scattering and nonresonant four-wave-mixing processes}

\author{Daniel L. Marks}
\author{Claudio Vinegoni}
\author{Jeremy S. Bredfeldt}
\author{Stephen A. Boppart}
  \altaffiliation[Also in ]{the Department of Bioengineering, the College of Engineering, and the College of Medicine, University of Illinois at Urbana-Champaign, Urbana IL 61801; \emph{boppart@uiuc.edu}}
\affiliation{
Beckman Institute for Advanced Science and Technology, Dept. of Electrical and Computer Engineering, University of Illinois at Urbana-Champaign, 405 N. Mathews, Urbana, IL 61801}

\date{\today}

\begin{abstract}
A major impediment of using Coherent Anti-Stokes Raman Scattering
to identify biological molecules is that the illumination levels
required to produce a measurable signal often also produce significant
nonresonant background from the medium, especially from water, that is not
specific to the resonance being investigated.  We present a method of
using nonlinear interferometry to measure the temporal shape of the
anti-Stokes signal to differentiate which components are 
resonant and nonresonant.  This method is easily adaptable to most
existing pulsed CARS illumination methods and should allow for distinguishing
resonant CARS when using higher energy pulses.  By examining the
differences between signals produced by acetone and water,
we show that the resonant and nonresonant signals can be clearly
differentiated.
\end{abstract}

                             % Classification Scheme.
%\keywords{Suggested keywords}%Use showkeys class option if keyword
                              %display desired
\maketitle

The combination of microscopy and Coherent Anti-Stokes Raman
Scattering (CARS)
processes~\cite{potma2002,cheng2002,dudovich2002,duncan1982} is a
promising tool to study the composition of biological tissues at
micrometer scales.  Like two-photon microscopy, CARS microscopy uses a
nonlinear interaction to produce a confined point response in the
medium.  However, CARS utilizes endogenous molecular resonances in the
tissue and does not require exogenous dyes or markers to be
introduced.  Because CARS consists of two stimulated Raman scattering
(SRS) processes, there is a quadratic dependence on the anti-Stokes
produced by CARS on the density of molecules with a target resonance.
Frequently the anti-Stokes signal is small because the desired target
molecule is present at a low concentration.  To compensate for this,
higher energy pulses are used.  However, at sufficiently high peak
power a large non-resonant four-wave-mixing component is generated.  If
the power of the anti-Stokes light is examined, one can not
distinguish the nonresonant signal from the desired resonant signal.
We utilize interferometry to distinguish the resonant CARS from
the nonresonant background based on the emission time by using the
interferometric time gate to reject the early-emitted nonresonant signal.
This is a simplification of the Nonlinear Interferometric Vibrational
Imaging (NIVI)~\cite{marks2004,vinegoni2004} method proposed earlier
that is more suited to integration with existing CARS pump/Stokes
pulse generation methods.

A typical CARS process consists of two SRS processes.  A molecule is
excited by two overlapped pulses, a pump pulse of frequency $ \omega_1
$ and a Stokes pulse of frequency $ \omega_2 $ separated by the
resonance frequency $ \Omega=\omega_1-\omega_2 $.  Some of this
excitation is converted to anti-Stokes radiation at frequency $
\omega_3=2\omega_1-\omega_2 $ by mixing with the pump.  The amount of
anti-Stokes radiation depends on the square of the intensity of the
pump pulse and linearly with the Stokes pulse.  Often the intensities
needed to produce a measurable CARS signal stimulate other nonlinear
nonresonant processes that do not depend on the presence of the target
molecule producing a non-negligible background signal, potentially
causing an erroneous concentration measurement.  Distinguishing the
processes that are resonant and therefore specific to a target
molecule and nonresonant processes is a significant limitation when
using CARS at high pulse energies.

Other means of distinguishing resonant CARS from nonresonant
signals have been explored.  Because phase-matching in bulk media
favors forward propagating anti-Stokes radiation, more backreflected
(epi-CARS) anti-Stokes is produced by smaller subwavelength sized
particles, and less so by the surrounding medium~\cite{volkmer2001}.
The production of resonant CARS can be favored by careful preparation
of the polarization and phase of the illumination~\cite{oron2002-4},
or selected by polarization~\cite{cheng2001-4}.  In addition, it is
possible to use a delayed probe pulse from the pump/Stokes pulse to
measure the resonant excitation if the probe pulse is differentiated
spatially or spectrally from the pump pulse~\cite{gershgoren2003}.
However, with the tight focusing required and in the presence of
highly scattering media, these methods may be less practical for
biological tissue.

To see how nonresonant signals and resonant CARS can be separated,
consider that nonresonant signals arise from four wave mixing
processes mediated by virtual states.  In nonresonant
four-wave-mixing, virtual states exist only where and when the pump
and Stokes pulses coincide, so that the anti-Stokes is only produced
at the instant they are overlapped.  Resonant CARS is produced because
a molecular vibrational, rotational, or electronic state is excited by
SRS.  This excitation persists after the excitation pulse ends, often
for a picosecond or more.  An analogous situation exists when
comparing the beating of a drum to the plucking of a guitar string.
Like a guitar string, the vibration of a molecular resonance decays
slowly, while a drum beat ends quickly after the impulse is over.  If
pump light continues to illuminate the molecule, the molecular
excitation can be converted by SRS to anti-Stokes radiation.  Because
the resonant excitation lasts much longer than the nonresonant
excitation, the anti-Stokes also lasts longer.  Thus anti-Stokes
radiation caused by resonant CARS continues to be emitted later than
the nonresonant signal.  With properly designed pulses, the resonant
and nonresonant signals can be clearly separated.

\begin{figure}
\includegraphics[bb=0 260 761 700,width=17 cm,clip=false]{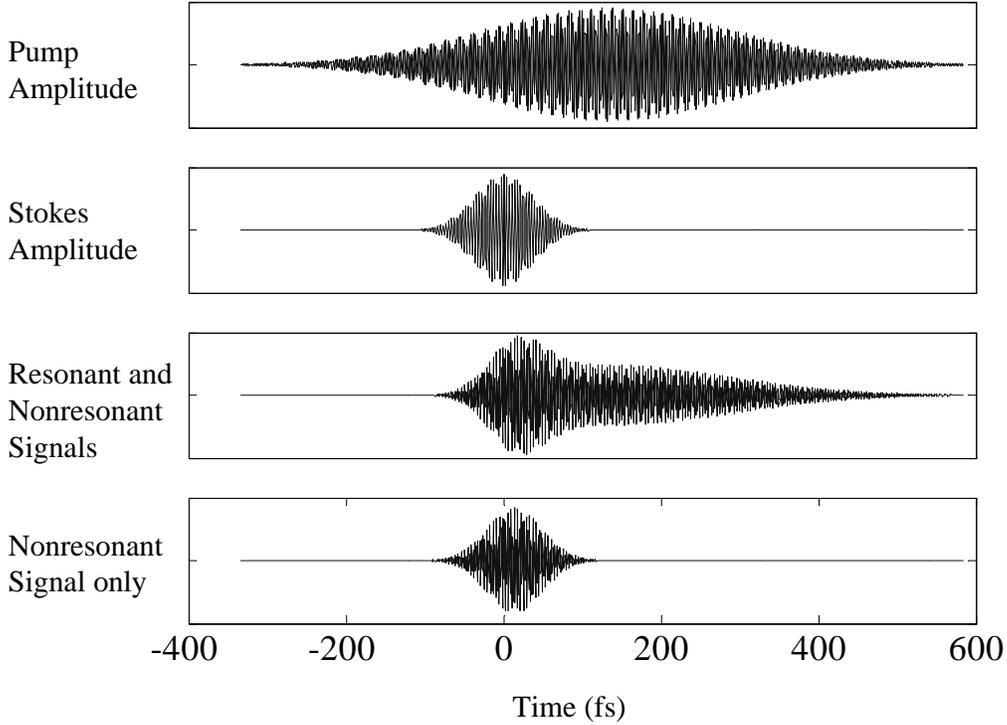}
\caption{A simulation of the pulse combination used to differentiate between resonant and nonresonant signals.  The pump/Stokes combination overlap to excite CARS, and the anti-Stokes would appear similar to the shown waveform for resonant or nonresonant media.}
\label{fig:fig1}
\end{figure}

Our approach is to prepare narrowband pump and Stokes pulses, but with
the pump pulse stretched out in time to be at least three times longer
than the Stokes pulse.  The shorter Stokes pulse coincides with the
leading edge of the pump pulse.  A simulation of this is shown in
Fig.~\ref{fig:fig1}. When the overlapped pulses arrive, the molecule
is excited by SRS.  At the same time, nonresonant four-wave-mixing is
emitted, overlapped with the Stokes pulse.  After the Stokes pulse
passes, so does the nonresonant signal.  However, the molecule remains
excited.  As the pump continues to arrive, the excitation is converted
to anti-Stokes radiation by SRS.  This produces a resonant anti-Stokes
signal similar to that shown in Fig.~\ref{fig:fig1}, which has a
resonant ``tail'' unlike the nonresonant anti-Stokes, which coincides
with the Stokes alone.  By delaying a reference pulse at the
anti-Stokes frequency until after the nonresonant signal has passed,
the reference can act as an interference gate to reject nonresonant
components.

Interferometric time gating is commonly used to characterize the shape
of ultrafast pulses~\cite{purchase1993,lepetit1995}.  These methods
typically work by interfering a reference pulse with a known electric
field amplitude with an unknown pulse to be characterized.  If the
reference pulse is short in time compared to the unknown pulse, then
interference between the two pulses only occurs over the interval of
the reference pulse.  By delaying the two pulses relative to each
other, the cross-correlation of the two pulses is measured.  A resonant
CARS signal has a much longer cross-correlation signal in time than a
nonresonant signal.  To obtain the needed short reference pulse,
nonresonant four-wave-mixing or other cascaded nonresonant nonlinear
processes stimulated by short pulses can be used.  In particular, a
reference pulse delayed until after the nonresonant signal arrives at
the photodetector prevents an interference signal from being obtained
from the nonresonant component.

\begin{figure}
\includegraphics[bb=50 450 861 700,width=20 cm,clip=false]{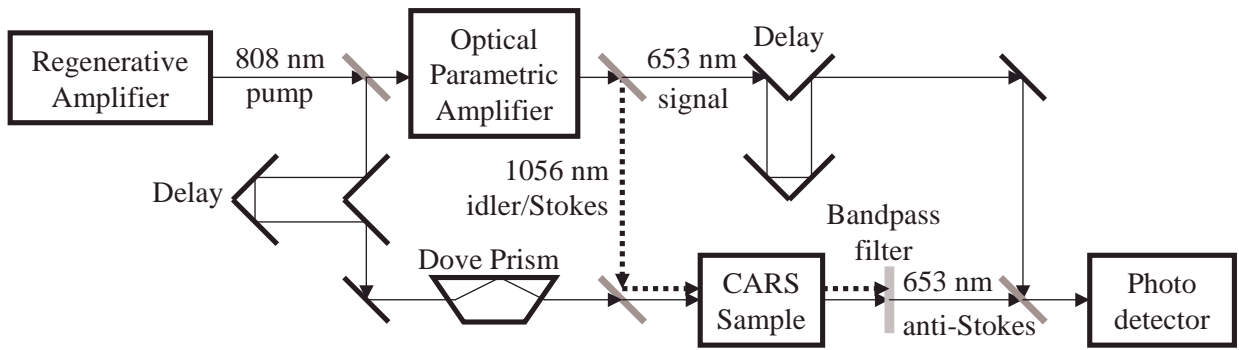}
\caption{Schematic of optical setup used to measure interferogram of resonant and nonresonant signals.}
\label{fig:fig2}
\end{figure}

To experimentally validate this idea, we used the setup of
Fig.~\ref{fig:fig2} to measure the interferograms of anti-Stokes light
produced by acetone and water.  Acetone has a Raman resonance at 2925
$\mbox{cm}^{-1}$ corresponding to the C-H stretch, while water does
not, containing only hydrogen and oxygen.  Water is of primary
interest because it is a ubiquitous and pernicious source of
nonresonant signal in biological tissues.  In the setup, a
regenerative amplifier (RegA 9000, Coherent, Inc. Santa Clara, CA)
emits pulses at 250 kHz repetition rate with 808 nm center wavelength
and 20 nm bandwidth.  These pulses are used both as the pump and 
also to seed a second-harmonic-generation optical parametric
amplfier (OPA) (OPA 9450, Coherent) which generates idler pulses with
1056 nm center wavelength and 20 nm bandwidth for use as a Stokes
pulse.  A 105 mm length BK7 glass Dove prism disperses the pump pulse
to approximately three times the length of the Stokes pulse.  The pump
pulse is delayed to arrive at a dichroic beamsplitter at the same time
as the Stokes pulse.  The pulses are overlapped and are focused into
the sample by a 30 mm focal length lens, which produces
anti-Stokes radiation centered at 653 nm.  The pump power at the
sample was 40 mW, while the Stokes was 2 mW, with sufficient peak
power to produce abundant resonant and nonresonant signals.  At the
same time, the signal pulse from the OPA, also at 653 nm, is used as
the reference pulse.  Because the signal pulse is produced by cascaded
nonresonant $\chi^{(2)}$ nonlinearities, it is short and nearly
transform-limited.  A Mach-Zehnder interferometer is used to combine
the reference pulse and the CARS signal.  The signals are attenuated
by neutral density filters by many orders of magnitude before they are
detected by a photomultiplier tube.  By scanning the relative delay
between the two signals, their interferometric cross-correlation was
measured.

\begin{figure}
\includegraphics[bb=50 160 861 650,angle=270,width=8 cm,clip=false]{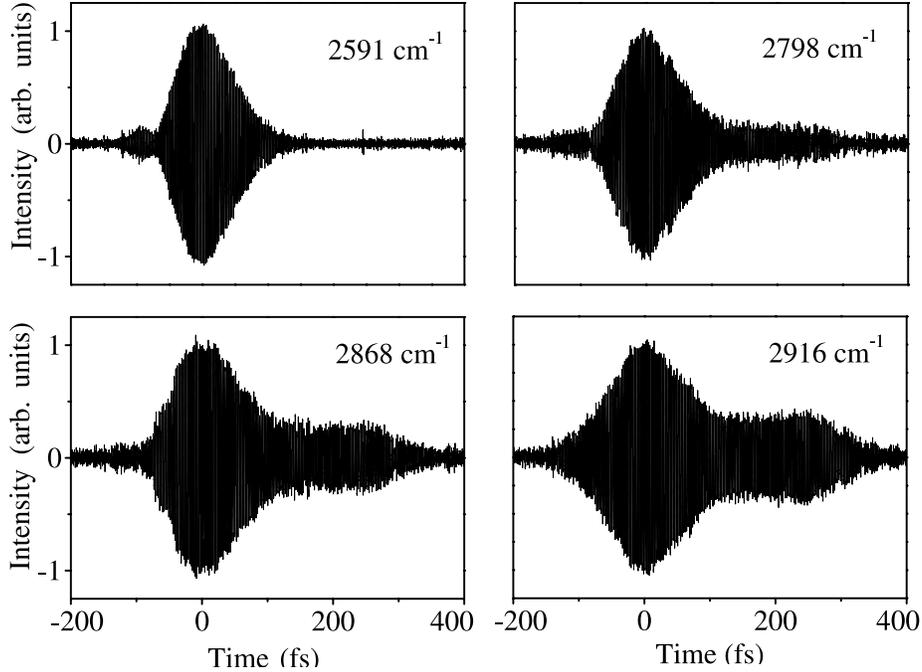}
\vspace{-4cm}
\caption{Interferogram of four-wave-mixing in acetone at various vibrational excitation frequencies.}
\label{fig:fig3}
\end{figure}

\begin{figure}
\includegraphics[bb=100 160 861 650,angle=270,width=8 cm,clip=false]{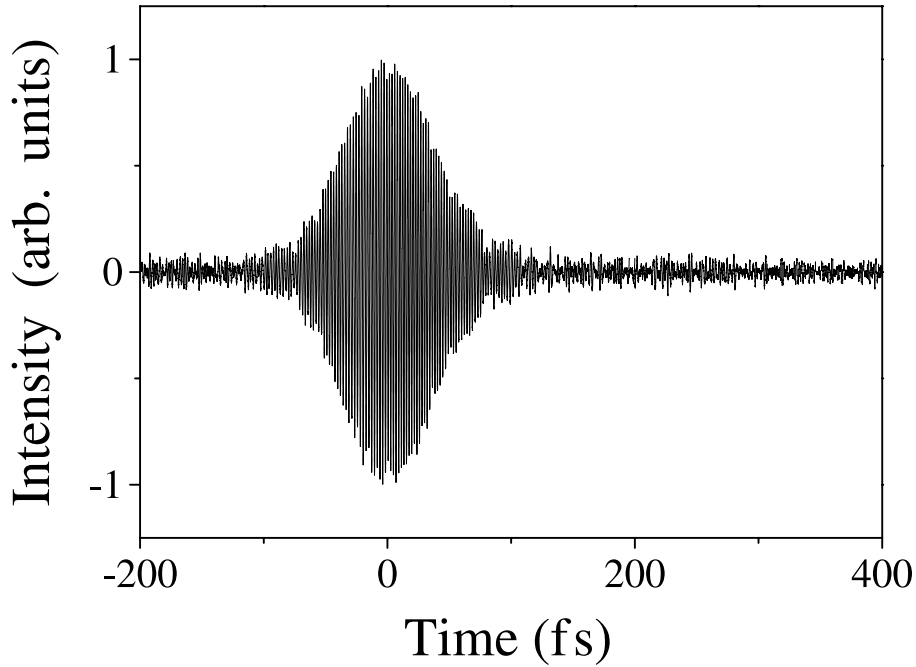}
\vspace{-4cm}
\caption{Interferogram of nonresonant four-wave-mixing in water.}
\label{fig:fig4}
\end{figure}

Fig.~\ref{fig:fig3} shows the interferograms measured from acetone.
As can clearly be seen, the interferograms agree qualitatively with
Fig.~\ref{fig:fig1}.  The acetone, having a persistent resonance,
generates a resonant anti-Stokes ``tail'' with a length limited not by
the lifetime of the resonance but by the length of the pump pulse,
because the pump is needed to produce anti-Stokes radiation.  As the
pump/Stokes frequency difference is tuned away from the resonance at
2925 $\mbox{cm}^{-1} $, the resonant ``tail'' disappears.  The tuning
resolution is limited by the broad Stokes bandwidth of approximately
150 $\mbox{cm}^{-1}$, which is much wider than the Raman
susceptibility linewidth.  On the other hand, Fig.~\ref{fig:fig4}
shows the interferogram from water, which is completely nonresonant at
2925 $\mbox{cm}^{-1} $.  The resonant and nonresonant signals are
discernible interferometrically despite the fact that the excitation
power used produced enough CARS light to be clearly seen by the
unaided eye scattered from white paper.

We have demonstrated a difference in the temporal evolution of
anti-Stokes pulses produced by nonresonant and resonant
four-wave-mixing processes.  This approach utilizes nonlinear
interferometry and appropriate reference and excitation pulses to
measure the tail of resonant CARS.  Such an approach will likely be
very useful in CARS microscopy and NIVI~\cite{marks2004,vinegoni2004}
to eliminate the nonresonant background signal in addition to the
other advantanges that interferometric detection can provide such as
heterodyne sensitivity and stray light rejection.

\section{Acknowlegements}
\label{sec:acknowledgement}
We acknowledge the scientific contributions and advice from
Martin Gruebele, Dana Dlott, Amy Wiedemann, and Barbara Kitchell from
the University of Illinois at Urbana-Champaign.  This research 
was supported in part by the National Aeronautics and Space
Administration (NAS2-02057), the National Institutes of Health
(National Cancer Institute), and the Beckman Institute for
Advanced Science and Technlogy.

\end{document}